\documentclass{iau} 
\usepackage{graphicx}

\title[Interferometry of class I methanol masers] %% give here short title %%
{Interferometry of class I methanol masers, statistics and the distance scale}

\author[M. A. Voronkov et al.]   %% give here short author list %%
{Maxim A. Voronkov$^{1,3}$, Shari L. Breen$^{2,3}$, Simon P. Ellingsen$^3$ \and
Christopher H. Jordan$^4$}

\affiliation{$^{1}$CSIRO Astronomy and Space Science, PO Box 76, Epping,
NSW 1710, Australia \\email: {\tt maxim.voronkov@csiro.au} \\[\affilskip]
$^2$Sydney Institute for Astronomy (SIfA), School of Physics, University of Sydney, Sydney, NSW 2006, Australia \\[\affilskip]
$^3$School of Mathematics and Physics, University of Tasmania, GPO Box
252-37, Hobart, Tas 7000, Australia \\[\affilskip]
$^4$International Centre for Radio Astronomy Research, Curtin University, Bentley, WA 6845, Australia}

\pubyear{2017}
\volume{336}  %% insert here IAU Symposium No.
\jname{Astrophysical Masers: Unlocking the Mysteries of the Universe}
\editors{A. Tarchi, M.J. Reid \& P. Castangia, eds.}
\begin{document}

\maketitle

\begin{abstract}
The Australia Telescope Compact Array (ATCA) participated in a number of survey programs to search for and image 
common class I methanol masers (at 36 and 44 GHz) with high angular resolution. In this paper, we discuss
spatial and velocity distributions revealed by these surveys. In particular, the number of maser regions is found to fall off 
exponentially with the linear distance from the associated young stellar object traced by the 6.7-GHz maser, 
and the scale of this distribution is 263$\pm$15 milliparsec.
Although this relationship still needs to be understood in the context of the broader field, it can be utilised to
estimate the distance using methanol masers only. This new technique has been analysed 
to understand its limitations and future potential. It turned out, it can be very successful to
resolve the ambiguity in kinematic distances, but, in the current form, is much less accurate 
(than the kinematic method) if used on its own. 
\end{abstract}

\firstsection % if your document starts with a section,
\section{Introduction}
Early studies of methanol masers empirically divided them into two classes (\cite[Batrla \etal\ 1987]{bat87}): class~I are typically
scattered around the presumed location of the young stellar object (YSO)  over an area often compared with the primary beam of
the 20-m class radio telescope, while class~II are compact at arcsecond resolution and pinpoint the location of the YSO. These differences 
were traced to collisional and radiative pumping, respectively (see \cite{vor14} and references therein for further information). 
The Australia Telescope Compact Array (ATCA) participated in a number of survey programs, both 
targeted and blind, to search for and image common class~I methanol masers (at 36 and 44 GHz) with 
high angular resolution (\cite[Voronkov \etal\ 2014]{vor14}; \cite[Jordan \etal\ 2015, 2017]{jor15,jor17}; and unpublished data 36-GHz data
from the Methanol Multibeam (MMB) follow-up project).
This resulted in a large sample of class~I methanol masers studied at sub-arcsecond resolution, and provided us the basis for statistical analysis.

\section{Velocity distribution}
Due to complexity of spatial and kinematic structure of class~I methanol masers, we decomposed emission into groups of Gaussian components
coincident within 3$\sigma$ in position and velocity and used those groups for statistical analysis instead of individual components 
(see \cite{vor14} for further details). Each component group was assigned an association with a 6.7-GHz maser (from the MMB project, see \cite{gre17} and references therein), 
if found within an arcminute (if there is more than one 6.7-GHz maser in the vicinity we used the nearest).  The middle of the velocity range spanned by the 
6.7-GHz maser emission is often used as a reference velocity for the kinematic distance estimates. The comparison between this mid-range velocity and velocity
of the associated class~I emission revealed that the difference is largely (ignoring a few high-velocity components) a Gaussian variate with the standard deviation of
3.65$\pm$0.05 and 3.32$\pm$0.07~km~s$^{-1}$ for the 36 and 44-GHz masers, respectively (\cite[Voronkov \etal\ 2014]{vor14}). 
The difference between the two standard deviations, if real, may be a consequence of preferential orientation with respect to the line of sight direction for 
the 36 and 44-GHz masers which belong to two different transition series  (\cite[Sobolev \etal\ 2007; see also the paper in this volume]{sob07}). The distribution of velocity offsets has
also a small but significant mean of $-$0.57~km~s$^{-1}$ (class~I masers are blue-shifted; uncertainties 
are 0.06 and 0.07~km~s$^{-1}$ for the 36 and 44-GHz masers, respectively). However, the velocities of class~I masers show a better agreement with that of the thermal
gas: the standard deviation of the velocity offsets with respect to CS is 1.5$\pm$0.1~km~s$^{-1}$ with insignificant mean offset (see \cite[Jordan \etal\ 2015, 2017]{jor15,jor17}).
Therefore, the velocity of class~I methanol masers seems to be a better tracer of the quiescent gas velocity than the velocity derived from the 6.7-GHz spectra.
This conclusion is in agreement with the earlier results of \cite{gre11} on the velocity distribution of 6.7-GHz masers.

\section{Spatial distribution}

Despite general scatter around the presumed YSO location, the smaller spatial separations of class~I maser emission are somewhat preferred with the 
region of influence of a YSO being less than 1~pc (see Fig. 29 of \cite[Kurtz \etal\ 2004]{kur04}).
The large statistics and availability of homogenous 6.7-GHz data through the MMB survey allowed us to probe this distribution in detail. It turned out
that the linear separations between class~I maser components and associated 6.7-GHz masers have an exponential distribution with
good accuracy (the scale is 263$\pm$15 milliparsec; see \cite[Voronkov \etal\ 2014]{vor14}). This relationship still needs to be understood in the context of the broader field.

\section{New method to estimate distances}

\begin{figure}[t]
% \vspace*{-2.0 cm}
\begin{center}
 \includegraphics[width=\linewidth]{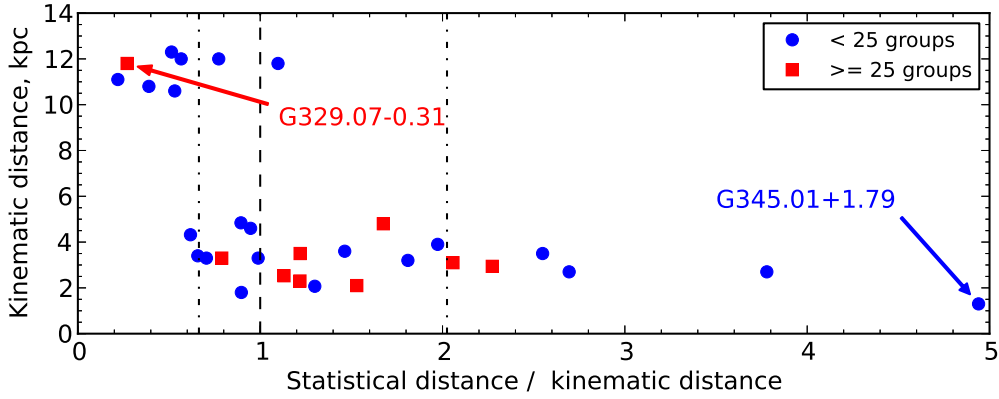} 
% \vspace*{-1.0 cm}
 \caption{Comparison of the distance estimate based on the empirical spatial distribution of class~I masers and the kinematic distance estimate based on the 6.7-GHz velocities. 
 Each point represents a single 6.7-GHz maser from the MMB catalogue associated with at least 15 class~I methanol maser component groups (sources with at least 25
 such groups are shown by (red) squares). The vertical  dash-dotted lines enclose the area where both methods agree (dashed vertical line represents the exact agreement) 
 with the 95\% confidence, under assumption that the new method is the only source of uncertainty and the class~I methanol maser has 15 component groups.}
   \label{distestplot}
\end{center}
\end{figure}

The empirical distribution of linear separations discussed in the previous section can, in principle, be turned into a statistical distance 
estimator from the measured angular separations. This method does not use velocities and, therefore, is applicable in the Galactic Centre direction where the kinematic method 
fails. Here we analyse the limitations of this new approach. First, the number of component groups for each individual source is rarely large. Therefore, it may be more practical to 
assume that the measured angular separation is the exponential variate for each individual source. Then, one can use the fact that, for the exponential 
distribution, the mean is an unbiased maximum likelyhood estimator of the scale parameter. And equating the measured mean angular separation ($\left<d_{arcsec}\right>$) to the expected linear scale of
263$\pm$15~milliparsecs (\cite[Voronkov \etal\ 2014]{vor14}), one gets the following recipe for the distance $D_{kpc}\approx54/\left<d_{arcsec}\right>$.
The numbers may be too low for a proper statistical test that the measured angular separations for a given source are indeed consistent with the exponential distribution. However, one could compute the standard
deviation and compare it with the mean as a cross-check: they are identical for the exponential distribution. Note, the uncertainty of the expected linear
scale quoted above has rather small contribution (about 5\%) to the total error budget which is dominated by the uncertainty of $\left<d_{arcsec}\right>$ in the case of small number statistics and, quite likely,
by the systematics discussed below.

Second, the distribution of linear separations, which this method is based upon, has been obtained for an ensamble of sources. Applying it to individual sources introduces additional systematics due to possible hidden
parameters and idiosyncrasies of the sources. In paricular, it is discussed by \cite{vor14} that more evolved sources show some tendency to be more spread out both spatially and in velocity. This evolutionary
trend has been completely ignored in the analysis above. In addition, the ATCA follow-up of the 44-GHz masers found in the MALT45 blind survey (\cite[Jordan \etal\ 2017]{jor17}) and the new (unpublished) 36-GHz 
ATCA survey targeting MMB 6.7-GHz masers, both revealed many simple sources. This is an additional evidence suggesting that the source sample studied by \cite{vor14} may not be representative of the whole population of class~I methanol masers (or, even, the subset of such masers located in high-mass star forming regions). Another important systematic factor is the association of class~I methanol masers with a range of phenomena producing shocked gas, not just outflows (see discussion in \cite[Voronkov \etal\ 2010; 2014]{vor10,vor14}). The spatial distribution of class~I masers conceivably depends on the exact mechanism, but details, or even the relative occurrence of different scenarios, are not clear at present. 

To examine the accuracy of such statistical distances in practical terms, we compared them with kinematic distances based on the middle of the velocity range of the associated 6.7-GHz maser for the whole sample
of \cite{vor14} where an association between class~I and class~II masers can be made and where the class~I maser emission was decomposed into at least 15 component groups. Each point in 
Fig.~\ref{distestplot} represents a single source. Statistical distances are expected to be more accurate for sources with large number of component groups which are shown as (red) squares. But, in general,
points are expected to be confined between two dash-dotted lines on the diagram (agreement with the kinematic method with 95\% confidence), or roughly a factor of 2 uncertainty in distances. 
However, it is evident from Fig.~\ref{distestplot} that there are many outliers with grossly overestimated distances for nearby sources and underestimated for sources beyond 10~kpc. Inspection of individual 
sources in the former category reveals two main reasons for overestimated distances: the sensitivity cutoff due to primary beam (e.g. the most extreme outlier point corresponding to G345.01$+$1.79 which
was found by \cite{vor14} to have emission beyond the half-power point of the primary beam) and questionable associations of the class~I maser emission with the 6.7-GHz masers in complex sources. In both
cases, the apparent spread of class~I maser emission is underestimated causing the method to overestimate the distance. However, it is worth noting that there are many relatively compact sources which have
less than 15 component groups. There seems to be a sysematic tendency to overestimate the distances to such sources, although some of them may indeed have been incorrectly assigned to the near kinematic
distance. For distant sources, extreme outliers (which this method tries to bring closer) are likely to be largely genuine cases of sources wrongly assigned to the far kinematic distance (see, for example, the discussion 
on G329.07$-$0.31 in \cite[Voronkov \etal\ 2014]{vor14}). Therefore, despite having a relatively low accuracy on its own, this method can be used to assist resolving the kinematic distance ambiguity. 
It can also provide an extra constraint for the Bayesian approach suggested by \cite[Reid \etal\ (2016; see also the paper in this volume)]{rei16}.

It is worth noting that this method can be extended to sources without a 6.7-GHz maser or other way to pinpoint the YSO location. The averaged position of all class~I maser emission is close to the location
of the 6.7-GHz maser for the majority of the sources in our sample (although there are several sources with pronounced asymmetry for which it is not the case) and can be used as a proxy. 
This is an additional source of systematic uncertainty, but may be acceptable as the statistical distances are the rough estimates anyway.


\begin{thebibliography}{}

\bibitem[Batrla \etal\ (1987)]{bat87}
{Batrla, W., Matthews, H.E., Menten, K.M., \& Walmsley, C.M.} 1987,
\textit{Nature}, 326, 49

\bibitem[Green \etal\ (2017)]{gre17}
{Green, J.A., Breen, S.L., Fuller, G.A., McClure-Griffiths, N.M., Ellingsen, S.P., Voronkov, M.A., Avison, A., Brooks, K., Burton, M.G., Chrysostomou, A., Cox, J., Diamond, P.J.,
Gray, M.D., Hoare, M.G., Masheder, M.R.W., Pestalozzi, M., Phillips, C., Quinn, L.J., Richards, A.M.S., Thompson, M.A., Walsh, A.J., Ward-Thompson, D., Wong-McSweeney, D., \& Yates, J.A.}
2017, \textit{MNRAS}, 469, 1383

\bibitem[Green \& McClure-Griffiths (2011)]{gre11}
{Green, J.A., \& McClure-Griffiths, N.M.} 2011, \textit{MNRAS}, 368, 1843

\bibitem[Jordan \etal\ (2017)]{jor17}
{Jordan, C.H., Walsh, A.J., Breen, S.L.,  Ellingsen, S.P., Voronkov, M.A., \& Hyland, L.J.} 2017,
\textit{MNRAS}, 471, 3915

\bibitem[Jordan \etal\ (2017)]{jor15}
{Jordan, C.H., Walsh, A.J., Lowe, V., Voronkov, M.A., Ellingsen, S.P., Breen, S.L., Purcell, C.R., Barnes, P., Burton, M.G., 
Cunningham, M.R., Hill, T., Jackson, J.M., Longmore, S.N., Peretto, N., \& Urquhart, J.S.} 2015,
\textit{MNRAS}, 448, 2344

\bibitem[Kurtz \etal\ (2004)]{kur04}
{Kurtz, S., Hofner, P., \& \' Alvarez, C.V.} 2004, \textit{ApJS}, 155, 149

\bibitem[Reid \etal\ (2016)]{rei16}
{Reid, M.J., Dame, T.M., Menten, K.M., \& Brunthaler, A.} 2016, \textit{ApJ}, 823, 77

\bibitem[Sobolev \etal\ (2007)]{sob07}
{Sobolev, A.M., Cragg, D.M., Ellingsen, S.P., Gaylard, M.J., Goedhart, S., Henkel, C., Kirsanova, M.S., Ostrovskii, A.B., Pankratova, N.V., Shelemei, O.V., 
van der Walt, D.J., Vasyunina, T.S., \& Voronkov, M.A.} 2007,  in: J.M. Chapman \& W.A. Baan (eds.), 
 \textit{Proc. IAU Symp. 242, Astrophysical Masers and their Environments} (Cambridge Univ. Press), p.\,81

\bibitem[Voronkov \etal\ (2014)]{vor14}
{Voronkov, M.A., Caswell, J.L., Ellingsen, S.P.,  Green, J.A., \& Breen, S.L.} 2014,
\textit{MNRAS}, 439, 2584

\bibitem[Voronkov \etal\ (2010)]{vor10}
{Voronkov, M.A., Caswell, J.L., Ellingsen, S.P., \&  Sobolev, A.M.} 2010,
\textit{MNRAS}, 405, 2471


\end{thebibliography}
\end{document}